\documentclass[conference, 9pt]{IEEEtran}
\IEEEoverridecommandlockouts
\ifCLASSINFOpdf
\else
   \usepackage[dvips]{graphicx}
\fi
\usepackage{url}

\usepackage{graphicx}
\usepackage{bm}
\usepackage{amssymb}
\usepackage{lipsum}
\usepackage{amsmath}
\usepackage[caption=false, font=footnotesize]{subfig}
\usepackage{enumitem}% http://ctan.org/pkg/enumitem
\usepackage{tikz}
\usepackage{lipsum}
\newcommand\copyrighttext{%
  \footnotesize © 2025 IEEE.  Personal use of this material is permitted.  Permission from IEEE must be obtained for all other uses, in any current or future media, including reprinting/republishing this material for advertising or promotional purposes, creating new collective works, for resale or redistribution to servers or lists, or reuse of any copyrighted component of this work in other works.}
\newcommand\copyrightnotice{%
\begin{tikzpicture}[remember picture,overlay]
\node[anchor=south,yshift=10pt] at (current page.south) {\fbox{\parbox{\dimexpr\textwidth-\fboxsep-\fboxrule\relax}{\copyrighttext}}};
\end{tikzpicture}%
}

\begin{document}
\intextsep 8pt 
\textfloatsep 2pt
\dbltextfloatsep 5pt
\belowcaptionskip 0pt

\title{\huge\textbf{Loudspeaker Beamforming to Enhance Speech Recognition Performance of Voice Driven Applications}}
\author{\IEEEauthorblockN{D. de Groot, B. Karslioglu,  O. Scharenborg, J. Martinez}\IEEEauthorblockA{\textit{Multimedia Computing Group, EEMCS, Delft University of Technology, The Netherlands}\\ \{d.c.c.j.degroot, j.a.martinezcastaneda, o.e.scharenborg\}@tudelft.nl, b.karslioglu@student.tudelft.nl}\vspace{-10pt}}

\maketitle

\pagestyle{plain}

\intextsep 8pt 
\textfloatsep 2pt
\dbltextfloatsep 5pt

\begin{abstract}
In this paper we propose a robust loudspeaker beamforming algorithm which is used to enhance the performance of voice driven applications in scenarios where the loudspeakers introduce the majority of the noise, e.g. when music is playing loudly. The loudspeaker beamformer modifies the loudspeaker playback signals to create a low-acoustic-energy region around the device that implements automatic speech recognition for a voice driven application (VDA). The algorithm utilises a distortion measure based on human auditory perception to limit the distortion perceived by human listeners. 
Simulations and real-world experiments show that the proposed loudspeaker beamformer improves the speech recognition performance in all tested scenarios. Moreover, the algorithm allows to further reduce the acoustic energy around the VDA device at the expense of reduced objective audio quality at the listener's location. 
\end{abstract}
\copyrightnotice
\begin{IEEEkeywords}
Spotforming, beamforming, speech recognition 
\end{IEEEkeywords}
\IEEEpeerreviewmaketitle

\vspace{-2pt}\section{Introduction}\label{01_Introduction}\vspace{-3pt}
\IEEEPARstart{A}{utomatic} speech recognition (ASR) is steadily being integrated into our daily lives. While ASRs can attain high accuracy in clean acoustic conditions, they are often deployed as part of voice driven applications (VDAs), such as voice assistants like \textit{Amazon Alexa}, \textit{Apple Siri} and \textit{Google Assistant}. The scenarios where VDAs should work include noisy and reverberant environments, e.g. living rooms and cars \cite{King2017, Vu2021, Minder2023}. These environments pose a significant challenge to perform ASR \cite{Umbach2021}, as the algorithms are usually not trained on different types of signal degradations but only on clean speech \cite{Whisper_AT, Whisper_flamingo}. In practice different types of speech enhancement techniques, e.g. denoising and microphone beamforming, are used as preprocessing steps to improve ASR performance \cite{Gannot2017, Kassir2022, Quan2024}. Yet, ASR performance in adverse acoustic conditions is not always satisfactory \cite{MISP2021}. This has prompted researchers to use a large amount of microphones \cite{Sadeghi2024}, or to consider different sensor modalities such as video \cite{Whisper_flamingo}, bone-conduction microphones \cite{Manzanillo2024} and radar \cite{Fan2023}. While effective, these systems rely on specialized (sometimes intrusive) hardware or to record video, which is not always feasible or desirable.
In this work, we propose a loudspeaker beamforming algorithm to improve ASR performance in VDA systems with multiple loudspeakers and where the loudspeaker signals are themselves a major source of interference. This is the case, for example, when the user is watching a movie in their home cinema and wants to pause the movie with a voice command or when the user likewise wants to pause music in their car. We are not the first to consider the loudspeaker modality to enhance ASR performance. In \cite{Miyabe2007}, a method combining sound field synthesis and acoustic echo cancellation (AEC) was proposed. Compared to our method, this method has more stringent requirements on the number of loudspeakers. Additionally, in order to perform AEC, it was assumed that the playback signals are available at the VDA. This assumption is not satisfied when only a low capacity link is available between the playback system and the VDA. Although AEC is not included in this work, our algorithm allows for its integration if the playback signals are available to the VDA.\looseness=-1

The proposed loudspeaker beamforming algorithm is based on a class of robust near-field microphone beamformers called microphone spotformers, in which the signal from some region-of-interest is amplified or attenuated \cite{Taseska2016, Martinez2015}. We adapt the microphone spotformer to create a ``loudspeaker spotformer'' (LSp). The proposed system modifies the loudspeaker playback signals such that a low-acoustic-energy region is formed around the VDA while maintaining a high reproduction quality with low perceived distortion around the user. Keeping the perceived distortion limited is challenging in practice. 
For this, an objective measure of human sound perception well-suited for optimisation is required. This is an active research topic in e.g. sound zone synthesis \cite{Lee2020, Koeijer2024}. We constrain the distortion introduced to the listener using a measure of human auditory masking \cite{Par2005}, as in \cite{Koeijer2024}. The proposed LSp algorithm assumes that the location of the loudspeakers, the user, and the VDA lie within regions that can be estimated or measured a priori. This is a reasonable assumption: in a car or a living room the locations of the loudspeakers, the user, and the VDA mostly remain within a priori established regions. Our proposed algorithm is evaluated through simulations and in a real-world scenario. The Matlab code implementing the loudspeaker spotformer can be found at \cite{Groot2024}. 
\looseness=-1
\vspace{-1pt}\section{Notation and signal model}\label{Sec2Not}\vspace{-2pt}
Consider a space in which $L$ loudspeakers are placed at positions $\mathbf{x}^{(l)}_\text{L}\in\mathbb{R}^3$, $l\in\left\{1, \ldots, L\right\}$. For each loudspeaker, we define $s^{(l)}_\text{ref}$ as the reference loudspeaker playback signal (no LSp algorithm applied). Likewise, $s^{(l)}_\text{L}$ is the loudspeaker playback signal when the LSp algorithm is applied. In the room there is a VDA equipped with a microphone array implementing microphone beamforming. The center of the array is located at $\mathbf{x}_\text{M}\in\mathbb{R}^3$. In practice, the locations of the VDA and loudspeakers can be measured by hand or estimated through signal processing techniques such as \cite{Heusdens2014}. There is a listener assumed to be approximately at a known location $\mathbf{x}_\text{u}\in\mathbb{R}^3$. We use the free-field acoustic transfer function from the loudspeakers to $P$ points in the neighbourhood of $\mathbf{x}_\text{u}$ as control points: $\mathbf{x}^{(p)}_\text{P}\in\mathbb{R}^3$, $p\in\{1, \ldots, P\}$. The room impulse response (RIR) from location $\mathbf{x}_s$ to location $\mathbf{x}_r$ is given by $h(\mathbf{x}_s, \mathbf{x}_r, t)$. Frequency-domain variables are indicated with a hat on top of the corresponding symbol. The room transfer function (RTF) is given by $\hat{h}(\mathbf{x}_s, \mathbf{x}_r, \omega)$, with $\omega=2\pi f$ and $f$ the frequency in hertz. We model each RTF as consisting of the direct path only, eliminating the need to estimate the highly varying RTF. Instead we rely on the robustness of the LSp algorithm. Thus, the RTF is given by \looseness=-1
\vspace{-0.1cm}
\begin{equation}
    \hat{h}(\mathbf{x}_s, \mathbf{x}_r, \omega) = \frac{e^{-j\omega ||\mathbf{x}_s-\mathbf{x}_r||_2/c}}{4\pi||\mathbf{x}_s-\mathbf{x}_r||_2},\vspace{-0.1cm}
    \label{eq1}
\end{equation}\vspace{-0.05cm}
with $j^2=-1$ and $c=342$ m/s the sound velocity \cite{Martinez2015, Ahrens2012}.

\vspace{-2pt}\section{Loudspeaker Spotforming for VDAs}\label{sec2_spot}
\begin{figure}[ht]
\vspace{-12pt}
    \centering
    \subfloat[Topview schematic of the setup\label{fig01aLoud}]{
        \includegraphics[width=0.48\columnwidth]{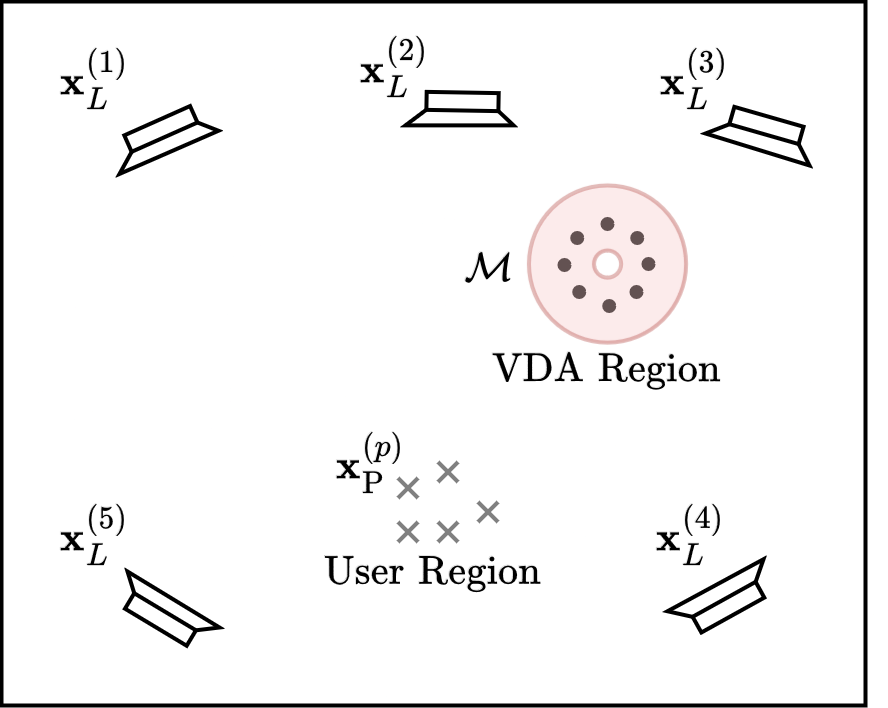}
    }
   \subfloat[Sideview photo of the real room\label{fig01bVA}]{
       \scalebox{1}[1]{  \frame{\includegraphics[width=0.464\columnwidth, trim={0cm 0cm 0cm 0cm}, clip]{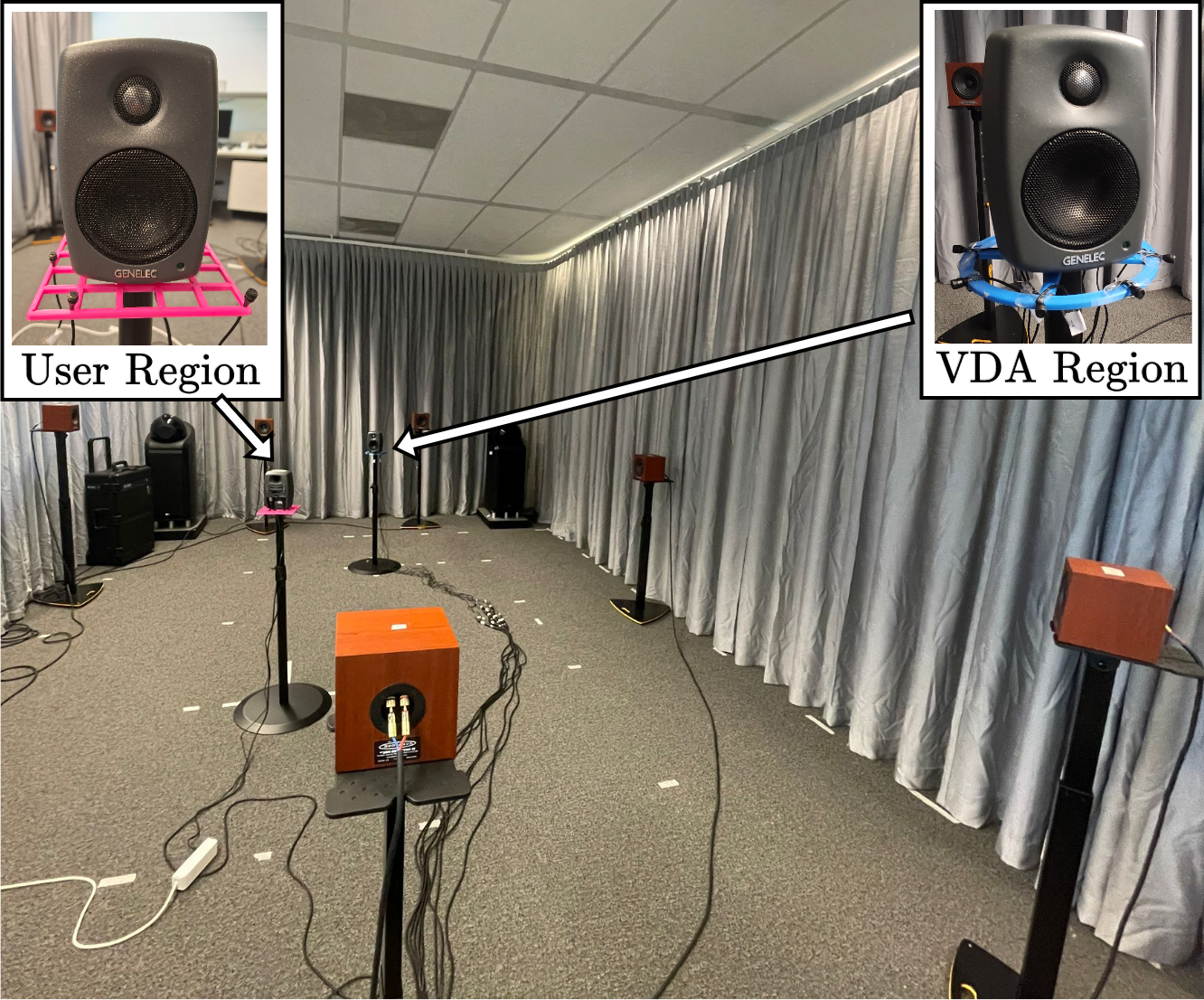}}}
    }
    \caption{The loudspeaker spotformer (LSp) setup. In (a), a topview schematic of the setup is shown. The LSp computes the loudspeaker playback signals which minimise the acoustic energy in region $\mathcal{M}$ around the microphones ($\bullet$) of the voice driven application (VDA). The control points ($\times$) are not physically placed but modelled within the region the user is expected to be listening. The algorithm limits the acoustic distortion at these points. In (b), a photo of the actual experiment setup is shown. A zoomed-in photo of our VDA implementation using a circular microphone array is shown in the top right corner. In the top-left corner, a zoom-in photo of the loudspeaker emulating the user in the experiments is shown. The microphones on the pink grid are used to evaluate the audio quality in Sec. \ref{sec4a}.\looseness=-1}\vspace{-4pt}
    \label{fig01}
\end{figure}

\vspace{-2pt}
In this section the setup used for our simulations and real-world test is introduced and the loudspeaker spotformer (LSp) is derived. In Fig. \ref{fig01} the setup is depicted. The left hand side (Fig. \ref{fig01aLoud}) provides a top-view schematic of the room with the position of the loudspeakers $\mathbf{x}_\text{L}^{(l)}$, the position of the VDA featuring a circular microphone array (region $\mathcal{M}$), and the position of the listener zone depicting the control points $\mathbf{x}_\text{P}^{(p)}$. To the right (Fig. \ref{fig01bVA}), a picture of the implemented setup in a real room is given. A zoom-in picture featuring our implementation of the VDA with microphone array is shown on the top right.  
In Fig. \ref{fig01aLoud} the microphones are depicted by dots ($\bullet$) and the control points by crosses ($\times$). Notice that region $\mathcal{M}$ (in red) is not a circle but a torus. Mathematically the torus reflects the spatial distribution of the circular microphone array more accurately.  \looseness=-1

In the following we compute the spatial covariance matrix $\mathbf{R}_\mathcal{M}$ describing the acoustic energy in region $\mathcal{M}$. The LSp algorithm is obtained by combining this covariance matrix with a distortion measure based on human auditory masking. \looseness=-1

\vspace{-3pt}\subsection{Spatial Covariance Matrix}\vspace{-2pt}
Define the loudspeaker-to-receiver transfer vector $\hat{\mathbf{v}}_\text{L}\in\mathbb{C}^L$ as 
\vspace{-0.1cm}\begin{equation}
    \hat{\mathbf{v}}_\text{L}(\mathbf{x}_r, \omega) = \begin{bmatrix} \hat{h}\left(\mathbf{x}_r, \mathbf{x}_\text{L}^{(1)}, \omega\right), \ldots,  \hat{h}\left(\mathbf{x}_r, \mathbf{x}_\text{L}^{(L)}, \omega\right) \end{bmatrix}^T\!\!\!,
\vspace{-0.1cm}\end{equation}
and the vector of frequency-domain loudspeaker signals $\hat{\mathbf{s}}_\text{L}$ as
\vspace{-0.1cm}\begin{equation}
    \hat{\mathbf{s}}_\text{L}(\omega) = \begin{bmatrix} \hat{s}_\text{L}^{(1)}(\omega), ..., \hat{s}_\text{L}^{(L)}(\omega) \end{bmatrix}^T\!\!\!.
\vspace{-0.1cm}\end{equation}
The audio received at $\mathbf{x}_r$ is given by 
$\hat{y}_r(\mathbf{x}_r, \omega) =  \hat{\mathbf{s}}_\text{L}^T(\omega)  \hat{\mathbf{v}}_\text{L}(\mathbf{x}_r, \omega)$. Now assume that $\mathbf{x}_r$ is a realisation of random vector $\mathbf{x}_\mathcal{M}$ with corresponding distribution $p_{\mathcal{M}}(\mathbf{x}_\mathcal{M})$, which can be interpreted as representing the probability of finding the microphones of the VDA over some spatial region. This spatial stochastic model is the key insight that provides the algorithm its robustness against errors in the positions. We return to this distribution later. Given random vector $\mathbf{x}_\mathcal{M}$, the signal received in region $\mathcal{M}$ is given by\looseness=-1
\vspace{-0.1cm}\begin{equation}
    y_\mathcal{M}(\mathbf{x}_\mathcal{M}, \omega) =  \hat{\mathbf{s}}_\text{L}^T(\omega)  \hat{\mathbf{v}}_\text{L}(\mathbf{x}_\mathcal{M}, \omega).
\vspace{-0.1cm}\end{equation}
The expected value of the acoustic energy in region $\mathcal{M}$ is
\vspace{-0.1cm}\begin{equation}
\begin{aligned}
    \mathbb{E}\left\{|y_\mathcal{M}(\mathbf{x}_\mathcal{M}, \omega)|^2\right\} &= 
    \hat{\mathbf{s}}^T_L(\omega)\mathbf{R}_{\mathcal{M}}(\omega) \hat{\mathbf{s}}^*_L(\omega)\\
    &=\left|\left|\mathbf{L}_\mathcal{M}^H(\omega)\hat{\mathbf{s}}^*_L(\omega)\right|\right|_2^2,
    \end{aligned}
      \label{eq9}
\vspace{-0.1cm}\end{equation}
with $\mathbb{E}\left\{ \hat{\mathbf{v}}_\text{L}(\mathbf{x}_\mathcal{M}, \omega)\hat{\mathbf{v}}^H_\text{L}(\mathbf{x}_\mathcal{M}, \omega)\right\}=\mathbf{R}_\mathcal{M}(\omega)=\mathbf{L}_\mathcal{M}(\omega)\mathbf{L}_\mathcal{M}^H(\omega)$. Matrix $\mathbf{L}_\mathcal{M}(\omega)$ follows from the Cholesky factorisation of positive-semidefinite matrix $\mathbf{R}_\mathcal{M}(\omega)$ \cite{Hayes1996},\cite[Cor. 7.2.9]{Horn2012}. 
The elements of $\mathbf{R}_{\mathcal{M}}(\omega)$ are found by spatially integrating over $p_\mathcal{M}$,
\vspace{-0.07cm}\begin{equation}
    \!\! \{\mathbf{R}_{\mathcal{M}}(\omega)\}_{ll'}\! = \!\! \int_{\mathbb{R}^3} \! \hat{h}\!\left(\mathbf{x}, \mathbf{x}_\text{L}^{(l)}, \omega\right) \! \hat{h}^* \! \! \left(\mathbf{x}, \mathbf{x}_\text{L}^{(l')}, \omega\right)p_{\mathcal{M}} \!\left(\mathbf{x}\right)d\mathbf{x}
    \label{eq10}.
\vspace{-0.13cm}\end{equation}
Matrix $\mathbf{R}_\mathcal{M}$ can be augmented to account for the contribution of late reverberation by making $\mathbf{R}_M=\mathbf{R}_\mathcal{M}+\mathbf{R}_\text{iso}$, where $\mathbf{R}_\text{iso}$ models the contribution of late reverberation as an isotropic acoustic field, the expression of which is known analytically \cite{Martinez2015}, \cite{Brandstein2001}. We return to distribution $p_\mathcal{M}$. This distribution can be interpreted as describing the probability of finding the circular microphone array of the VDA within some spatial region. We express the distribution in cylindrical coordinates with radius $r\in[0,\infty)$, azimuthal angle $\theta\in[0, 2\pi)$ and height $z\in\mathbb{R}$. Due to the circular microphone array, the distribution is chosen to resemble a torus centered at $\mathbf{x}_\text{M}$. This is achieved by using a uniform distribution in $\theta$ and Gaussian distributions in $z$ and $r$. The latter is truncated such that $r\geq 0$ \cite[p. 156]{Johnson1994}. The total distribution then is \vspace{-0.15cm}\looseness=-1
\begin{equation}
    p_\mathcal{M}(r, \theta, z) = \frac{C_r}{4\pi^2\sigma_r\sigma_z}e^{-\frac{1}{2}\left(\frac{(z-\mu_z)^2}{\sigma_z^2}+\frac{(r-\mu_r)^2}{\sigma_r^2}\right)},
    \label{eq7dis}
\vspace{-0.07cm}\end{equation}
with $r\geq0$, $\theta\in[0,2\pi)$, $C_r$ a normalising constant following from the truncated distribution and $\mu_r$ and $\mu_z$ the radius and height of the microphone array. The standard deviations $\sigma_r$ and $\sigma_z$ can be chosen based on the assumed measurement accuracy.\looseness=-1 

\vspace{-1pt}\subsection{The loudspeaker spotformer}\label{secSpot}\vspace{-2pt}
Using the covariance matrix we can minimise the acoustic energy in region $\mathcal{M}$. However, to avoid audible artefacts and to prevent the algorithm to converge to a trivial solution (i.e. turning all the loudspeakers off), it is important to constrain the algorithm in a smart way. We use a computationally inexpensive perceptual distortion measure based on tonal masking \cite{Par2005} which has been used in sound zone synthesis \cite{Koeijer2024}. The measure operates on short-time frames and predicts if a distortion $\hat{\bm{\epsilon}}$ is noticeable in the presence of an audible sound (called the {masker}) $\hat{\mathbf{s}}$. The distortion measure is given by $D(\hat{\mathbf{s}}, \hat{\bm{\epsilon}}) = ||\mathbf{P}_s\hat{\bm{\epsilon}}||_2^2$ \cite{Par2005}, with $\mathbf{P}_s$ is a diagonal matrix based on $\hat{\mathbf{s}}$ defining a frequency dependent weighting of the distortion. See \cite{Par2005} for details on the computation of $\mathbf{P}_s$. We define each masker as the audio signal at each control point when all loudspeakers play their reference signal $s^{(l)}_\text{ref}$. The disturbance is the deviation from the masker. \looseness=-1

From here on we switch the notation to discrete time-domain and discrete frequency-domain for the actual implementation of the algorithm. We use a frame-length $N_t$ and zero-padding length $N_\text{pad}$ so that the frequency-domain frame length is $N=N_t+N_\text{pad}$. Define the time-domain playback signal vector $\mathbf{s}_L^{(l)}=\begin{bmatrix} s^{(l)}(t_1) & \cdots & s^{(l)}(t_{N_t}) \end{bmatrix}^T\!\!$. The matrix of playback signals $\mathbf{S}_\text{L}$ is
\vspace{-0.1cm}\begin{equation}
    \mathbf{S}_\text{L} = \begin{bmatrix} 
        \mathbf{s}_\text{L}^{(1)} & \cdots & \mathbf{s}_\text{L}^{(L)}
    \end{bmatrix}\in\mathbb{R}^{N_t \times L}.
      \label{eq13}
\vspace{-0.2cm}\end{equation}
The frequency-domain matrix of playback signals $\hat{\mathbf{S}}_L$ is
\vspace{-0.1cm}\begin{equation}
    \hat{\mathbf{S}}_\text{L} = \mathbf{F}
    \begin{bmatrix}
        \mathbf{I} &
        \mathbf{0}
    \end{bmatrix}^T\mathbf{S}_\text{L}
    =\mathbf{W}\mathbf{S}_\text{L}\in\mathbb{C}^{N \times L}
      \label{eq14}
\vspace{-0.1cm}\end{equation}
where $\mathbf{F}$ is the $N\times N$ discrete Fourier transform matrix, $\mathbf{I}$ the $N_t\times N_t$ identity matrix and $\mathbf{0}$ an $N_t\times N_\text{pad}$ all-zero matrix. Therefore, $\mathbf{W}\in\mathbb{C}^{N\times N_t}$ defines a zero-padded discrete Fourier transform. The time- and frequency-domain matrices of reference signals $\mathbf{S}_\text{ref}$ and $\hat{\mathbf{S}}_\text{ref}$ are defined in the same manner.
Define the loudspeaker-to-control-point transfer matrix $\mathbf{V}_\text{P}^{(l,p)}\in\mathbb{C}^{N\times N}$ as 
\vspace{-0.1cm}\begin{equation}
    \!\mathbf{V}_\text{P}^{(l,p)} \!\! = \! \text{diag}\!\begin{bmatrix} \hat{h}\!\left(\mathbf{x}_\text{L}^{(l)}\!, \mathbf{x}_\text{P}^{(p)}\!\!, \omega_0\right)\!, \ldots,  \hat{h}\!\left(\mathbf{x}_\text{L}^{(l)}\!, \mathbf{x}_\text{P}^{(p)}\!\!, \omega_{N-1}\right) \end{bmatrix}^T\!\!\!\!\!.
      \label{eq12}
\vspace{-0.1cm}\end{equation}
Let $\{\mathbf{A}\}_k$ be an operator that extracts the $k^\text{th}$ column of $\mathbf{A}$. The masker $\hat{\mathbf{s}}_\text{ref}^{(p)}$ for control point $p$ is given by 
\vspace{-0.1cm}\begin{equation}
    \hat{\mathbf{s}}_\text{ref}^{(p)} = \textstyle\sum_{l=1}^{L}\mathbf{V}_\text{P}^{(l,p)}\{\hat{\mathbf{S}}_\text{ref}\}_l,
    \label{eq15}
\vspace{-0.1cm}\end{equation}
with corresponding masking-matrix $\mathbf{P}_\text{ref}^{(p)}$ as defined in \cite{Par2005}. 
The LSp is given as a convex optimisation problem \cite{Boyd2009} where the distortion at the control points is constrained. Using \eqref{eq9}, \eqref{eq14},  \eqref{eq12}, and \eqref{eq15} this gives \looseness=-1
\vspace{-0.05cm}\begin{equation}\setlength{\jot}{0pt}
    \begin{aligned}
        \min_{\hat{\mathbf{S}}_L} & \textstyle\sum_{k=0}^{N-1} \alpha(\omega_k) \left|\left|\mathbf{L}_M^H(\omega_k) \{\hat{\mathbf{S}}_L^T\}_k\right|\right|_2\\
        \text{s.t.}  &\quad  \hat{\mathbf{S}}_L = \mathbf{W}\mathbf{S}_L,  \quad \mathbf{S}_L\in\mathbb{R}^{N_t\times L},\\
        &\quad \left|\left|\mathbf{P}_\text{ref}^{(p)}\left( \textstyle\sum_{l=1}^{L}\mathbf{V}_\text{P}^{(l,p)}\{\hat{\mathbf{S}}_\text{L}\}_l-\hat{\mathbf{s}}_\text{ref}^{(p)}\right)\right|\right|_2^2 \leq d,\quad\forall p,
    \end{aligned}
    \label{eq16}
\vspace{-0.1cm}\end{equation}
where the $k$ in $\omega_k$ represents the $k^\text{th}$ frequency bin, $d$ is a parameter controlling the maximum allowable distortion calibrated to $d=1$ when the distortion is just noticeable and $\alpha(\omega_k)$ a user-defined weighting term (see Section \ref{sec4aSAP}).\looseness=-1

\section{Experimental Results}\label{Sec3_EXP}\vspace{-1pt}
In this section, the performance of the proposed LSp is evaluated. To the best of our knowledge, this is the first time loudspeaker beamforming is used to enhance voice driven applications (VDAs). Our main experiments are given in Sec. \ref{sec4b}, where we test how the addition of LSp affects the ASR performance of a VDA. The word error rate (WER) and word information lost (WIL) are used as metrics for testing the ASR performance.
Before this, in Sec. \ref{sec4a}, we analyse the ability of the LSp to create a low-acoustic-energy region around the VDA while keeping a high reproduction quality with low perceived distortion at the user position. This is important, as our main hypothesis is that if this is correctly achieved an increase in ASR speech recognition performance follows straightforwardly. \looseness=-1 
The experimental setup is described in Sec. \ref{sec4aSAP}. All experiments are carried out through simulations and by implementation in a real-world environment. The latter is important to test the claimed robustness and actual applicability of the algorithm. \looseness=-1 

\vspace{-1pt}\subsection{Experimental setup and algorithm parameterisation}\label{sec4aSAP}
\begin{figure}[ht]
\vspace{-3pt}
    \centering
 {\includegraphics[width=\columnwidth, trim={2.2cm 0.6cm 1cm 0.5cm}, clip]{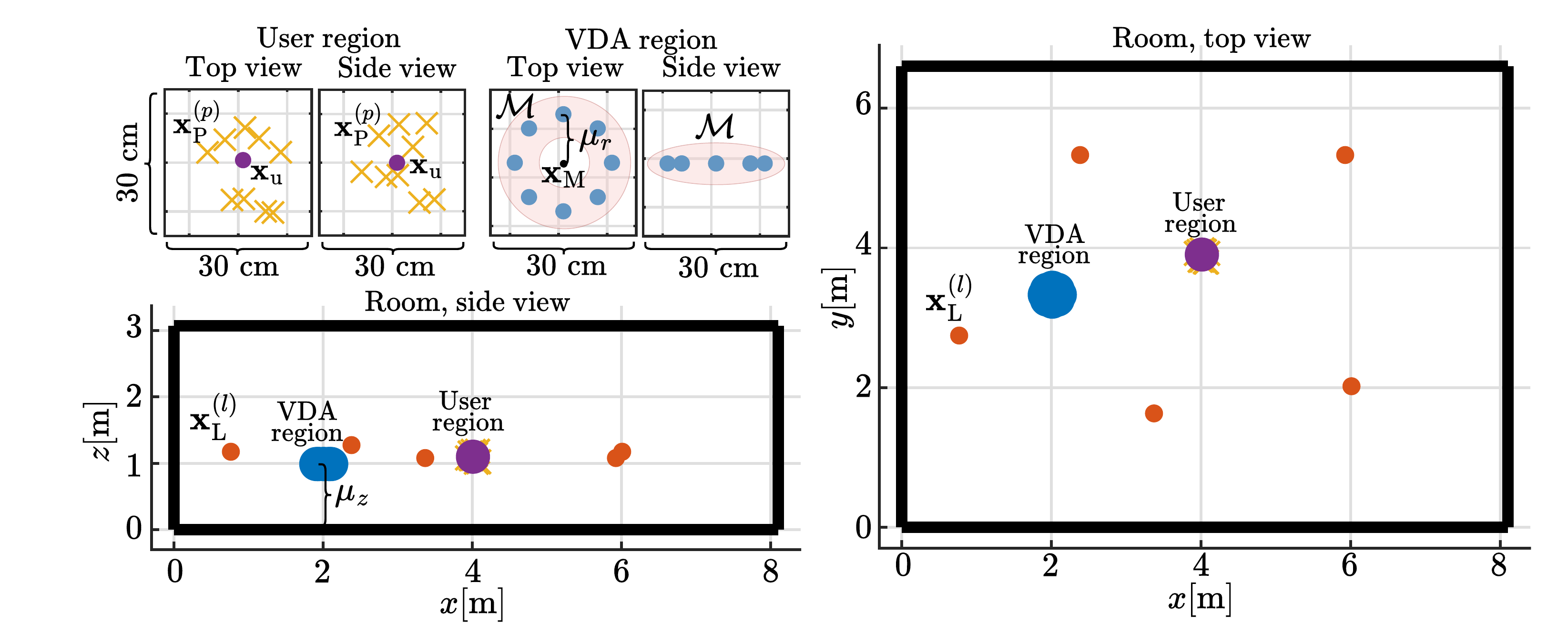}}
    \vspace{-20pt}
    \caption{A schematic view of the setup used in both simulations and real-world experiments. Zoom-ins on the user region (including control points $\mathbf{x}_\text{P}^{(p)}$and the user location $\mathbf{x}_\text{u}$) and on the VDA are provided. The microphone array of the VDA has a radius $\mu_r=0.1$ cm and a height $\mu_z=0.99$ m. Region $\mathcal{M}$ is placed approximately on top of these microphones and is centered at the VDA location $\mathbf{x}_\text{M}$. The loudspeakers are located at positions $\mathbf{x}_\text{L}^{(l)}$. }
    \label{fig02}
\end{figure}

\vspace{-1pt}
As mentioned in Sec. \ref{01_Introduction} our target application consists of systems with loudspeakers and a VDA like a voice assistant, where the loudspeaker signals are themselves a major source of interference for ASR. We choose our setup as to roughly represent a home cinema in a living room were the users are watching a movie and one would want to give a voice command to e.g. pause the movie.
In the experiments we use five loudspeakers and consider a user giving voice commands. Current VDA devices are often equipped with a microphone array to enhance ASR performance. Therefore, our VDA implementation has a circular microphone array with eight microphones. The experiment setup used in both simulations and real-world experiments is shown in Fig. \ref{fig02}. For a photo of the real room and the VDA, see Fig. \ref{fig01bVA}. \vspace{-4pt}

\vspace{-5pt}In our target applications, the system would be in cases used to play music, or to make a phone-call (or video-call). For our tests we choose the loudspeaker playback signals to be segments of: an instrumental rock song, a female-voiced jazz song, a male-voiced pop song, a female-voiced speech signal and a male-voiced speech signal. Thus, the test signals consist of 40\% speech, 40\% vocal music and 20\% instrumental music. To evaluate the ability of the LSp to create a low-acoustic-energy region around the VDA we use white Gaussian noise due to its broadband characteristic. Since the spectral content of speech is limited above 7 kHz \cite{Deng2003} and to reduce computational complexity, a sample-rate of 16 kHz is used. \looseness=-1

Our LSp algorithm is given by optimisation problem \eqref{eq16} and is solved for the loudspeaker playback signals using CVX \cite{Grant2014}. Eq. \eqref{eq16} is formed through equations \eqref{eq9}, \eqref{eq10}, \eqref{eq7dis}, \eqref{eq14},  \eqref{eq12}, and \eqref{eq15}.   We now describe the parameters used for these equations. For \eqref{eq7dis}, which mathematically describes region $\mathcal{M}$ (see Fig. \ref{fig02}) we use $\mu_r=0.1$ m (the radius of the microphone array), $\mu_z=0.99$ m (the height of the microphone array) and $3\sigma_r=3\sigma_z=0.095$ m. The number of modelled control points is $P=9$  placed in space sampled from a normal distribution centered at the user ($\mathbf{x}_\text{u}$) with a standard deviation of $3\sigma=0.2$ m. Since most important information for recognising speech starts at a frequency of approximately 100 Hz ($\omega=200\pi$ rad/s) \cite[Ch. 7.2.1]{Deng2003}, the regularisation parameter $\alpha(\omega_k)=0$ in \eqref{eq16} equals $0$ if $|\omega_k|\leq200\pi$ rad/s and $\alpha(\omega_k)=1$ otherwise. As in \cite{Jeannerot2022}, we set the corresponding elements of matrix $\mathbf{P}_\text{ref}^{(p)}$ to high values (to 100 times the maximum of the inverse masking curve of the segment), encouraging the algorithm to keep the frequency content in this range unchanged. Audio is processed in segments of 16 ms (zero-padded to 32 ms) using 16 ms square-root Hanning window for analysis and synthesis with 8 ms hop length \cite{Smith2011}. We solve \eqref{eq10} numerically. We select a Clenshaw-Curtis quadrature method for a good compromise between numerical complexity and accuracy \cite{Press1992}. 
For the real-world experiments, an RME Fireface UFX+ audio interface is used. We emulated the VDA in the real-world experiments using a microphone array and a loudspeaker (top-right of Fig. \ref{fig01bVA}). The microphones are AKG C417 PP lavalier microphones and the loudspeaker is a Genelec 8010AP studio monitor. The loudspeakers of the playback system are Auratone 5C Super Sound Cubes. The `user' giving the voice commands is emulated by another Genelec 8010AP studio monitor (top-left of Fig. \ref{fig01bVA}). The eight microphones in the pink grid are used as validation points in the evaluation (see Section \ref{sec4a}). To incorporate measurement inaccuracies in the simulated results we used 100 runs in which the loudspeaker locations deviate from their expected locations with a standard deviation of 5 cm and where the microphones of the VDA are rotated on the azimuthal plane with a standard deviation of $5^\circ$. \looseness=-1

\vspace{-1pt}\subsection{Stand-alone loudspeaker spotformer performance test}\label{sec4a}\vspace{-1pt}
In this section we analyse the ability of the LSp to create
a low-acoustic-energy region around the VDA while keeping a low perceived distortion at the user position. This is evaluated in the following scenarios: \looseness-1
\begin{enumerate}[noitemsep,topsep=0pt]
    \item An anechoic simulated scenario.
    \item A simulated scenario with reverberation added using the mirror image source method (MISM) \cite{Allen1979} as implemented by Habets \cite{Habets2010}. The reverberation time $T_{60}\approx220$ ms.  
    \item A real room with a reverberation time $T_{60}\approx 220$ ms. 
\end{enumerate}

\begin{figure}[ht]
    \vspace{-12pt}
    \centering
     \subfloat[Objective audio quality \label{fig03aVISQOL}\vspace{-2pt}]{
        {\includegraphics[width=0.46\columnwidth, trim={1.2cm 0.8cm 0.67cm 0.6cm}, clip]{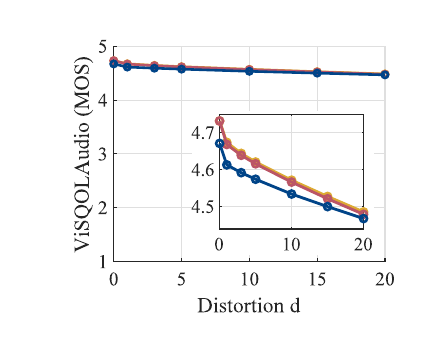}}
    }
   \subfloat[Reduction in received energy \label{fig03b}\vspace{-2pt}]{
        {\includegraphics[width=0.46\columnwidth]{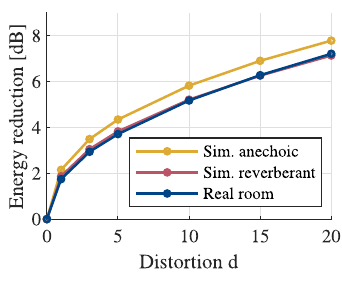}}
    }
    \caption{The results of the objective speech quality metric (including zoom-in) at the validation points given as mean-opinion-score (MOS) (a) and the reduction in received energy at the microphones (b) as function of distortion parameter $d$. In (a) the presented results are  averages over the different test signals and validation points. In (b) the results of the simulations are averaged over the 100 runs and the eight microphones of the array. The results of the real-world scenario are averaged over the eight microphones of the array. \looseness=-1\vspace{-2pt}}
    \label{fig03}
\end{figure}

To evaluate the distortion introduced by the LSp to the listener, the objective audio quality around the listener is measured as a function of the distortion parameter $d$ using ViSQOLAudio \cite{Hines2015, Chinen2020}. The loudspeaker signals used are our standard set of test signals, as named at the end of Section \ref{sec4aSAP}. The 100 simulated validation points (not to be confused with the control points $\mathbf{x}_\text{P}^{(p)}$) were sampled from a normal distribution centered at the user $\mathbf{x}_\text{u}$ with a standard deviation of 20 cm. The 32 validation points in the real-world experiments were taken using the $18\times18$ cm pink grid (top-left Fig. \ref{fig01bVA}) around the listener location at four different heights in steps of 5 cm. The results are plotted in Fig. \ref{fig03aVISQOL} as a mean opinion score (MOS) as a function of the allowed distortion $d$, where a MOS of 1 means bad perceived audio quality and 5 means excellent \cite{ITU800}.\looseness=-1

To evaluate the LSp energy-reduction performance, the achieved reduction in received energy at the microphones of the VDA is measured. This is done by comparing the energy received when the modified signal $s_\text{L}^{(l)}$ is playing (LSp turned on) with respect to when the reference signal $s_\text{ref}^{(l)}$ is playing (LSp turned off). The reference playback signal is white Gaussian noise. The results are shown in Fig. \ref{fig03b}. \looseness=-1 
In Fig. \ref{fig03aVISQOL} it can be seen that in all scenarios the objective audio quality at the listener location reduces only slightly as the allowed distortion increases: for all experiments the MOS score remains between `excellent' (5) and `good' (4). To put this in  context, for our chosen maximum distortion $d=20$, an acoustic energy reduction of minimum 7 dB (which is about a quarter of the original energy) is achieved (see Fig. \ref{fig03b}). And even in this case the MOS value remains high at about 4.4. 
Note that in both experiments depicted in Fig. \ref{fig03} the performance of the algorithm in the real-world scenario is not far from the performance in simulations. Considering that in a real implementation there are several sources of inaccuracy and aspects that are not modelled these results suggest robustness of the proposed algorithm. Moreover the curve's monotonic decrease in Fig. \ref{fig03aVISQOL} and Fig. \ref{fig03b} suggest that a trade-off between energy-reduction and perceived quality can be predictably and reliably achieved using parameter $d$. \looseness=-1
 
\subsection{Influence of loudspeaker spotforming on ASR performance}\label{sec4b}
In the previous section it is shown that the LSp can achieve a significant acoustic energy reduction around the VDA while keeping a good objective audio quality at the listener position. In this section we test our main hypothesis, namely how the addition of our novel LSp affects the ASR performance of voice driven applications. A distortion value $d=5$ is used. This is determined empirically: by listening to the output of the LSp ourselves, we found that up to $d=5$ the distortion remained mostly unnoticeable. \looseness=-1

We evaluate the ASR performance via the word error rate (WER) and the word information lost (WIL) \cite{Morris2004}. WIL ranges from 0 to 1 (lower is better) and is included because the voiced test signals introduce a large number of insertions, biasing the WER results.
The results are reported for different signal-to-interferer ratios (SIRs), where the ``signal" is the user voice command and the ``interferer" is the combined (summed) loudspeaker signals at the reference microphone of the VDA.
The ASR is Whisper Medium \cite{Whisper}. As user voice commands we use excerpts from the LibriSpeech database \cite{LibriSpeech} with a length of at most 6 seconds. For the interferer signals we use the test signals described in Sec. \ref{sec4a}. In total 2.1 hours playback runtime is made per SIR using different combinations of voice commands and interferer signals. The total time of voice commands is 0.5 hours. \looseness=-1

\begin{figure}[ht]
\vspace{-10pt}
\centering
   \subfloat{
        {\includegraphics[height=2.826cm,  trim={0cm 0.78cm 0.25cm 0.7cm}, clip]{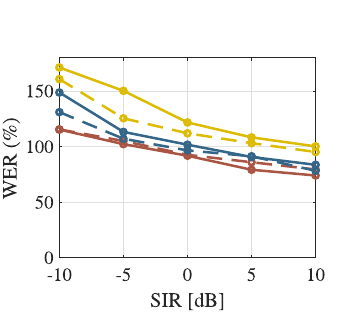}}
    }\hspace{-6pt}
   \subfloat{
        {\includegraphics[height=2.826cm,  trim={0.65cm 0.78cm 0.0cm 0.7cm}, clip]{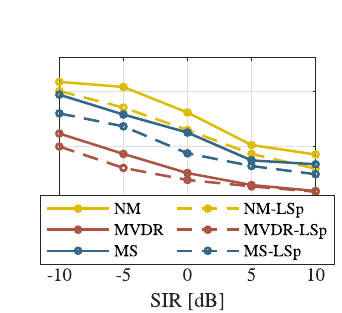}}
    }\\ \setcounter{subfigure}{0}
   \vspace{-10pt}
    \subfloat[Simulated reverberant\label{fig05b}\vspace{-2pt}]{
        {\includegraphics[height=3.411cm, trim={0cm 0cm 0.25cm 0.7cm}, clip]{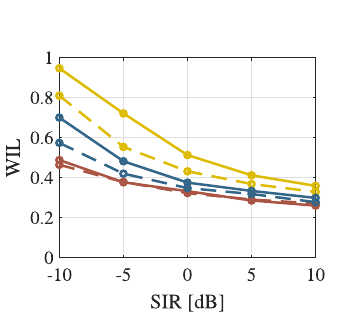}}
    }\hspace{-6pt}
    \subfloat[Real room\label{fig05c}\vspace{-2pt}]{
        {\includegraphics[height=3.411cm, trim={0.65cm 0cm 0.0cm 0.7cm}, clip]{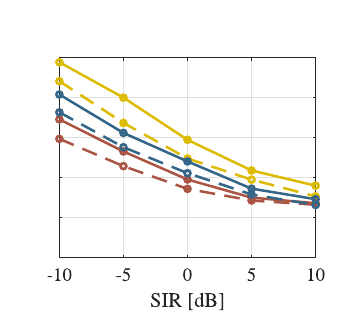}}
    }
    \caption{The results for word error rate (WER, upper row) and word information lost (WIL, lower row) as function of SIR in simulated reverberant conditions (a) and in the real room (b), where lower is better, for the three microphone beamformers (indicated by the different colours) and with (dashed line) and without (solid line) loudspeaker spotformer (LSp). The shown results are the average results over the different test signals and voice commands. The high WER is due to the interfering speech signals, outliers were removed.\looseness=-1}
    \label{fig05WER}
\end{figure}

Current VDAs include microphone arrays to improve the audio quality \cite{Ji2024}. This is done using microphone beamforming algorithms. The LSp algorithm is thus evaluated by: \looseness=-1
\begin{enumerate}[noitemsep,topsep=0pt]
    \item Tests excluding microphone beamforming. In practice, this would mean no microphone array is present in the VDA. Therefore we only use the signal from the nearest microphone (NM) to the user. \looseness=-1
    \item Tests including microphone beamforming. We select two beamformers:  (a) MVDR beamforming with oracle (perfect) voice activity detection \cite{Gannot2017, Martinez2015} and (b) the microphone spotformer (MS) proposed in \cite{Martinez2015}. MVDR is a classic algorithm used in many applications, therefore worth including and the LSp is inspired on the class of microphone spotformers. It is therefore interesting to test these algorithms working together. 
\end{enumerate}   
The results for the simulated reverberant environment and the real-world experiments are shown in Fig. \ref{fig05WER}. 

 Fig. \ref{fig05WER} shows that the average WER is predominantly more than 80\% in all experiments. This highlights the difficulty of speech recognition when there are interfering speech signals at high volumes, such as when pausing a movie with dialogue using a voice command. Furthermore, the results show that using the LSp consistently improves the ASR performance on average, particularly in low SIR conditions. This confirms our main hypothesis; namely that placing the VDA in the low-acoustic-energy region created by the LSp improves ASR performance. Importantly this also holds in the real room, suggesting the robustness and applicability of the proposed algorithm.

\section{Conclusion}\label{05_conclusion}
We proposed a robust loudspeaker spotformer to improve speech recognition performance of voice driven applications (VDAs) in noisy backgrounds, e.g. when music is playing loudly. The proposed algorithm creates a low-acoustic-energy region around the VDA while limiting the distortion introduced to the user of the VDA. Our main hypothesis was that if the VDA is placed in the low-acoustic-energy region the automatic speech recognition (ASR) performance increases. Our experiments showed that the low-acoustic-energy region can be created while keeping the objective audio quality high. Furthermore, the algorithm provides a parametric trade-off between acoustic energy reduction and reproduction quality. We confirmed our main hypothesis, namely the loudspeaker spotformer can improve ASR performance. We verified this in a real room, suggesting the robustness of our algorithm. A limitation of this work is the computational complexity involved in solving optimisation problem \eqref{eq16}. Therefore an efficient or analytic solution for \eqref{eq16} is a topic for future work. Additionally, the perceived audio quality at the user location should be confirmed through subjective tests. Lastly, other aspects of reproduction quality such as preservation of interaural-time and -level differences should be tested. \looseness=-1 

%\section*{References}
%\newpage
{}

\end{document}